\documentclass[pra,superscriptaddress,showpacs,twocolumn,10pt]{revtex4}

\usepackage{amssymb}
\usepackage{amsmath}
\usepackage{graphicx}

\newtheorem{theorem}{Theorem}
\newtheorem{lemma}{Lemma}

\begin{document}

\title{Unambiguous discrimination among quantum operations}
\author{Guoming Wang}
\email{wgm00@mails.tsinghua.edu.cn}
\author{Mingsheng Ying}
\email{yingmsh@tsinghua.edu.cn}

\affiliation{State Key Laboratory of Intelligent Technology and Systems,
Department of Computer Science and Technology, Tsinghua University, Beijing,
China, 100084}

\date{\today}

\begin{abstract}

We address the problem of unambiguous discrimination among a given
set of quantum operations. The necessary and sufficient condition
for them to be unambiguously distinguishable is derived in the
cases of single use and multiple uses respectively. For the latter
case we explicitly construct the input states and corresponding
measurements that accomplish the task. It is also found that the
introduction of entanglement can improve the discrimination.

\end{abstract}


\maketitle

The study of open quantum systems is an important subject in the
fields of quantum control and quantum information theory. Such
systems can be generally described in the quantum operations
formalism. Specifically, the behavior of an open system can be
represented by a linear, completely positive, trace-preserving map
$\cal E$, which is written in the Kraus operator-sum form
\cite{KR83}
\begin{equation}
{\cal E}(\rho)=\sum_{k}E_{k}\rho E_{k}^{\dagger},
\end{equation}
where $E_{k}$ are linear operators and satisfy the completeness
condition $\sum_{k}{E_{k}^{\dagger}E_{k}}=I$ in order to preserve
the trace of $\rho$.

Now the following problem naturally arises: if we are given a
quantum mechanical black box that performs one of the operations
${\cal E}_1,\dots,{\cal E}_n$, how can we identify which one it
really performs?  A natural idea is to input a probe state to the
black box and then distinguish between the possible outputs.
Moreover, if the black box can be accessed multiple times, we can
repeat the procedure and collectively discriminate the output
states of multiple uses.

Since this method relies on the discrimination of output states,
it is necessary to review the results about quantum state
discrimination at first. It is well known that a set of quantum
states can be perfectly distinguished if and only if that they are
orthogonal to each other. How to distinguish a set of
nonorthogonal quantum states in an optimal way according to some
criterion has become an important problem and received a lot of
attention in the past years \cite{BH04}. Two strategies are widely
used for this task. One is called minimum-error discrimination,
which allows mistakes but minimizes the probability of giving an
erroneous result. The other one, named unambiguous discrimination
\cite{IV87,DI88,PE88,CH01,SB02,EL03,RS03,TB03,RL03,CH04,ES04,FD04,HB04,
MS04,MH05,BH05,HB05,BH052,RL05},may fail for a nonzero
probability, but when it succeeds the result is absolutely right.
To be specific, in the task of unambiguous discrimination among
$\rho_1,\dots,\rho_n$, we need to construct a positive-operator
valued measure(POVM) comprising $n+1$ elements
$\Pi_0,\Pi_1,\dots,\Pi_n$ such that the measurement outcome $i$
correctly indicates $\rho_i$ for any $i=1,\dots,n$ and the outcome
$0$ leads to no conclusion. Unambiguous discrimination cannot be
applied to arbitrary set of states. It is proved \cite{FD04} that
the states $\rho_1,\dots,\rho_n$ can be unambiguously
discriminated if and only if for any $i=1,\dots,n$,
$supp(\{\rho_1,\dots,\rho_n\})\neq supp(\{\rho_j:j \neq i\})$, or
equivalently, $supp(\rho_i)\not\subseteq supp(\{\rho_j:j \neq
i\})$, where $supp(\rho_i)$ is the support of $\rho_i$, and the
support of a set of density operators is defined to be the sum of
each one's support. \cite{SS}.

The two strategies above can both be extended to the case of
quantum operations. However, neither of them is well studied so
far. Most previous work was directed to the special cases of
unitary operations \cite{CP00,AC01,CH05} and Pauli channels
\cite{AS05,SA052}. Some measures were also defined to quantify the
distinguishability of general quantum operations
\cite{GL05,BD05,YA05}. Only recently the problem of minimum-error
discrimination between two general quantum operations was
addressed by Sacchi \cite{SA05}.

In this paper we consider the problem of unambiguous
discrimination among a given set of quantum operations. The
necessary and sufficient condition for them to be unambiguously
distinguishable is derived in the cases of single use and multiple
uses respectively. For the latter case we explicitly give a
strategy. It is also found that the introduction of entanglement
can improve the discrimination.

We firstly consider the simple case when the black box can be
accessed only once. The problem can be formulated as follows: if
the possible quantum operations are ${\cal E}_1,\dots,{\cal E}_n$,
can we find a state $\rho$ in the input Hilbert space $\cal H$
such that ${\cal E}_1(\rho),\dots, {\cal E}_n(\rho)$ are
unambiguously distinguishable? More generally, we can introduce an
ancilla and make the main system and ancilla entangled to improve
our discrimination. Denoting the ancillary Hilbert space by ${\cal
H}_a$, our task is to find a state $\rho$ in the composite space
${\cal H}\otimes{\cal H}_a$ such that $({\cal E}_1\otimes {\cal
I})(\rho),\dots, ({\cal E}_n \otimes{\cal I})(\rho)$ can be
unambiguously discriminated, where $\cal I$ is the identity
operator acting on the space ${\cal H}_a$. If such ancillary space
and input state exist, we say that ${\cal E}_1,\dots,{\cal E}_n$
are unambiguously distinguishable by a single use.

Since any mixed input state can be purified by appending a
reference system which can be viewed as a part of the ancilla, we
just need to consider pure input states. Furthermore we know from
Schmidt decomposition that any ancillary space has a subspace of
dimension at most $dim({\cal H})$ that really matters in the
discrimination. Thus in the following we only need to consider the
pure states of the composite space ${\cal H} \otimes {\cal H}_a$
where $dim({\cal H}_a)=dim({\cal H})$.

Now we define the support of a quantum operation $\cal E$, denoted
by $supp({\cal E})$, to be the span of its Kraus operators
$\{E_k\}$, i.e.
\begin{equation}
supp({\cal E}) \equiv span\{E_k\} \equiv \{\sum_{k}{\lambda_k E_k
: \lambda_k \in {\mathbb C}\}}.
\end{equation}
It is proved that every two sets of Kraus operators describing the
same quantum operation $\cal E$ can be related to each other by a
unitary transformation \cite{NC00}. This fact indicates that our
concept $supp({\cal E})$ is independent of the specific choice of
Kraus operators so it is well-defined.

Furthermore, we define support of a set of quantum operations
$\{{\cal E}_1,\dots,{\cal E}_n\}$, denoted by $supp(\{{\cal
E}_1,\dots,{\cal E}_n\})$, to be the sum of every operation's
support, i.e.
\begin{equation}
supp(\{{\cal E}_1,\dots,{\cal E}_n\}) \equiv
\sum_{k=1}^{n}{supp({\cal E}_k)}.
\end{equation}

It is found out that the above concept of support of quantum
operations plays a very similar role like the support of quantum
states in determining the possibility of unambiguous
discrimination, as the following theorem indicates:

\begin{theorem}
The quantum operations ${\cal E}_1, \dots, {\cal E}_n$ can be
unambiguously discriminated by a single use if and only if for any
$i=1,\dots,n$, $supp({\cal E}_i) \not \subseteq supp(S_i)$, where
$S_i=\{{\cal E}_j:j\neq i\}$. \label{theorem_single}
\end{theorem}

{\it Proof.} Suppose ${\cal E}_i$ has Kraus operators
$\{E_{i}^{k}:k=1,..,n_i\}$. If the operation ${\cal E}_i \otimes
{\cal I}$ acts on the input $|\psi\rangle \in {\cal H}\otimes{\cal
H}_a$, the corresponding output is
\begin{equation} ({\cal E}_i\otimes {\cal
I})(|\psi\rangle\langle\psi|)= \sum_{k=1}^{n_i}{(E_{i}^{k}\otimes
I)|\psi\rangle\langle\psi| (E_{i}^{k}\otimes I)^{\dagger}}.
\end{equation}
It follows that its support is given by
\begin{equation}
supp(({\cal E}_i\otimes {\cal
I})(|\psi\rangle\langle\psi|))=span\{(E_{i}^{k}\otimes
I)|\psi\rangle:k=1,\dots,n_i\}. \label{equ:supp:span}
\end{equation}

If there exists a operation ${\cal E}_i$ satisfying $supp({\cal
E}_i)\subseteq supp(S_i)$, then we have that each $E_{i}^{k}$ can
be written as the linear combination of the operators
$\{E_{j}^{l}:j\neq i\}$. So for any input state $|\psi\rangle$,
$(E_{i}^{k}\otimes I)|\psi\rangle$ can also be written as the
linear combination of $\{(E_{j}^{l}\otimes I)|\psi\rangle:j\neq
i\}$. By Eq.(\ref{equ:supp:span}), this indicates
\begin{equation}
supp({\cal E}_i\otimes {\cal I})(|\psi\rangle\langle\psi|))
\subseteq supp(\{({\cal E}_j\otimes {\cal
I})(|\psi\rangle\langle\psi|):j\neq i\}).
\end{equation}
So it is impossible to unambiguously distinguish the output
$({\cal E}_i\otimes {\cal I})(|\psi\rangle\langle\psi|))$ from
other possible outputs $({\cal E}_j\otimes {\cal
I})(|\psi\rangle\langle\psi|)(j\neq i)$. Therefore we cannot
unambiguously distinguish the operation ${\cal E}_i$ from the
others.

The proof of the converse needs a constructive method. Now we
assume that the ${\cal E}_1, \dots, {\cal E}_n$ fulfill the given
condition. Let $|\psi\rangle$ be arbitrary pure state with full
Schmidt number, i.e.
\begin{equation}
|\psi\rangle=\sum_{t=1}^{d}{\alpha_{t}|t\rangle|t_a\rangle},
\label{equ:state:shc}
\end{equation}
where $\alpha_{t}>0$,$t=1,\dots,d$, $\{|t\rangle:t=1,\dots,d\}$
and $\{|t_a\rangle:t=1,\dots,d\}$ are orthonormal bases for ${\cal
H}$ and ${\cal H}_a$ respectively. We now prove that the set of
output states $\{({\cal E}_i\otimes {\cal
I})(|\psi\rangle\langle\psi|):i=1,\dots,n\}$ are unambiguously
distinguishable. Otherwise, there exists one operation ${\cal
E}_i$ satisfying
\begin{equation}
supp(({\cal E}_i\otimes {\cal I})(|\psi\rangle\langle\psi|))
\subseteq supp(\{({\cal E}_j\otimes {\cal
I})(|\psi\rangle\langle\psi|):j\neq i\}),
\end{equation}
Hence, by Eq.(\ref{equ:supp:span}) we know that there exist
coefficients $\{\lambda_{jl}^{k}\}$ such that for any
$k=1,\dots,n_i$,
\begin{equation}
(E_{i}^{k}\otimes I)|\psi\rangle = \sum_{j\neq i,
l=1,\dots,n_j}\lambda_{jl}^{k} (E_{j}^{l}\otimes I)|\psi\rangle.
\label{equ:linear:lin}
\end{equation}
Taking Eq.(\ref{equ:state:shc}) into Eq.(\ref{equ:linear:lin}), we
have
\begin{equation}
\sum_{t=1}^{d}\alpha_{t}E_{i}^{k}|t\rangle|t_a\rangle =
\sum_{t=1}^{d}\alpha_{t}\sum_{j\neq
i,l=1,\dots,n_j}\lambda_{jl}^{k}E_{j}^{l}|t\rangle|t_a\rangle.
\end{equation}
Since $\{|t_a\rangle\}$ are orthogonal to each other and
$\alpha_t>0$, we have that for any $|t\rangle$,
\begin{equation}
E_{i}^{k}|t\rangle = \sum_{j\neq
i,l=1,\dots,n_j}\lambda_{jl}^{k}E_{j}^{l}|t\rangle,
\end{equation}
which implies that
\begin{equation}
E_{i}^{k}=\sum_{j\neq i,l=1,\dots,n_j}\lambda_{jl}^{k}E_{j}^{l}.
\end{equation}
So we obtain $supp({\cal E}_i) \subseteq supp(S_i)$, which
contradicts the assumption. \hfill $\blacksquare$

It should be noted that from the proof above, any entangled pure
state with full Schmidt number can be used as input to universally
distinguish arbitrary set of quantum operations that fulfill the
condition in Theorem \ref{theorem_single}.

A corollary of the Theorem \ref{theorem_single} is that two
quantum operations ${\cal E}_1$ and ${\cal E}_2$ can be
unambiguously discriminated by a single use if and only if
$supp({\cal E}_1) \not \subseteq supp({\cal E}_2)$ and $supp({\cal
E}_2) \not \subseteq supp({\cal E}_1)$.

For the case in which we are not allowed to introduce any
ancillary system, a similar argument shows that the condition
presented in Theorem \ref{theorem_single} is still necessary. But
in general it is not sufficient. Let us consider the following
example. Suppose we are going to discriminate two Pauli channels,
the bit-flip channel and the phase-flip channel, whose Kraus
operators are $\{\sqrt{p}I,\sqrt{1-p}X\}$ and
$\{\sqrt{q}I,\sqrt{1-q}Z\}$ respectively, i.e.
\begin{eqnarray}
{\cal E}_1(\rho) & = & p\rho+(1-p)X\rho X,\\
{\cal E}_2(\rho) & = & q\rho+(1-q)Z\rho Z.
\end{eqnarray}
It is impossible to unambiguously distinguish these two channels
without use of ancilla. To see this, we notice that two qubit
states can be unambiguously discriminated only if they are both
pure. However, the inputs that make the output of the bit-flip
channel pure are
$|\pm\rangle=\frac{1}{\sqrt{2}}(|0\rangle\pm|1\rangle)$ but for
the phase-flip channel such inputs are $\{|0\rangle,|1\rangle\}$.
So for any input state $|\psi\rangle$, the outputs ${\cal
E}_1(|\psi\rangle\langle\psi|)$ and ${\cal
E}_2(|\psi\rangle\langle\psi|)$ cannot both be pure and thus they
are not unambiguous distinguishable. On the other hand, from
Theorem \ref{theorem_single} it is easy to see that when using
ancillay systems these two channels can be unambiguously
discriminated by a single use.

From the above example we see that the introduction of
entanglement between the main system and ancilla not only
increases the success probability but also in fact changes the
possibility of unambiguous discrimination between quantum
operations. It should be noted that for minimum-error
discrimination the use of entangled input can also increase the
efficiency \cite{SA05}.

Now we consider a more complicated case in which the black box can
be accessed multiple times. We say that quantum operations ${\cal
E}_1,\dots,{\cal E}_n$ can be unambiguously discriminated by $N$
uses if there exist an ancillary space ${\cal H}_a$ and a state
$\rho$ in the composite space ${\cal H}^{\otimes N}\otimes {\cal
H}_a$ such that $({\cal E}_1^{\otimes N}\otimes {\cal
I})(\rho),\dots,({\cal E}_n^{\otimes N}\otimes {\cal I})(\rho)$
are unambiguously distinguishable, where $\cal I$ is the identity
operators acting on ${\cal H}_a$. Still we only need to focus on
the case of $\rho = |\psi\rangle\langle\psi|$ where $|\psi\rangle
\in {\cal H}^{\otimes N}\otimes {\cal H}_a$ and $dim({\cal
H}_a)=dim({\cal H}^{\otimes N})$.

One may think that when a repeated use of the black box is
allowed, we can use quantum process tomography \cite{CN97,PC97} to
identify it. However, this method depends on the statistical data
of measurement outcomes and thus requires considerable effort. Our
method based on state discrimination accesses the black box a much
smaller number of times, so it is more efficient.

Now it is necessary to review some results about unambiguous
discrimination between quantum states with multiple copies because
they play a central role in the proof of our following theorem. It
is well known that a set of pure states can be unambiguously
discriminated if and only if they are linearly independent.
However, in \cite{CH01} Chefles found that even linearly dependent
pure states can be unambiguously discriminated if sufficient many
copies of them are distinguished collectively. A bound on the
number of copies needed was also obtained: for any $n$ distinct
pure states $|\psi_1\rangle,\dots,|\psi_n\rangle$ in a
$d$-dimensional space, $|\psi_1\rangle^{\otimes
c},\dots,|\psi_n\rangle^{\otimes c}$ can always be unambiguously
discriminated if $c \ge n-d+1$. Here we find a similar result for
mixed states.

\begin{lemma}
If the mixed quantum states $\rho_1,\dots,\rho_n$ satisfy that for
any $i\neq j$, $supp(\rho_i)\not \subseteq supp(\rho_j)$, then
$\rho_1^{\otimes n},\dots,\rho_n^{\otimes n}$ can be unambiguously
discriminated. Otherwise, for arbitrary $N \ge 1$,
$\rho_1^{\otimes N},\dots,\rho_n^{\otimes N}$ are not
unambiguously distinguishable.
\label{lemma_state}
\end{lemma}

{\it Proof.} If there exist two states $\rho_i$ and $\rho_j$ such
that $supp(\rho_i) \subseteq supp(\rho_j)$, then we have that for
any $N \ge 1$, $supp(\rho_i^{\otimes N}) \subseteq
supp(\rho_j^{\otimes N})$, so it is impossible to unambiguously
distinguish between the state $\rho_i^{\otimes N}$ and
$\rho_j^{\otimes N}$.

Now suppose that for any $i\neq j$, $supp(\rho_i)\not \subseteq
supp(\rho_j)$. Then we know that for any $i\neq j$, $supp(\rho_i)$
is not fully orthogonal to $ker(\rho_j)$, where $ker(\rho_j)$ is
the kernel of $\rho_j$ \cite{K}. Denoting the projection operators
onto $supp(\rho_j)$ and $ker(\rho_j)$ by $P_j,Q_j$ respectively,
then we obtain
\begin{equation}
tr(Q_{j}\rho_{i})>0,
\label{equ:mul:state}
\end{equation}
for any $i\neq j$.

If we have $n$ copies of the unknown state, numbered from $1$ to
$n$, we perform the projective measurement $\{P_{i}, Q_{i}\}$ on
the $i$th copy individually. Equivalently, a projective
measurement consisting of all the projection operators
$\{\Xi_{1}\otimes \Xi_{2} \dots \otimes \Xi_{n}\}$ is performed on
the n-fold copies, where for any $k=1,\dots,n$, $\Xi_{k}=P_{k}$ or
$Q_{k}$.

Consider the probability of getting the measurement outcome
corresponding to $Q_{1} \otimes \dots Q_{i-1} \otimes P_{i}
\otimes Q_{i+1}\otimes\dots\otimes Q_{n}$. If the unknown state is
$\rho_i^{\otimes n}$, the probability is
\begin{equation}
\begin{array}{l}
tr((Q_{1} \otimes \dots Q_{i-1}\otimes P_{i} \otimes Q_{i+1} \otimes \dots \otimes Q_{n}) \rho_i^{\otimes n}) \\
=tr(P_{i}\rho_i) \prod\limits_{j \neq i}{tr(Q_{j}\rho_i)} \\
>0,
\end{array}
\end{equation}
where the last inequality is derived from Eq.
(\ref{equ:mul:state}) and $tr(P_{i}\rho_i)=1$.

Otherwise, if the unknown state is $\rho_{j}^{\otimes n}$ for some
$j \neq i$, the probability is
\begin{equation}
\begin{array}{l}
tr((Q_{1} \otimes \dots Q_{i-1}\otimes P_{i} \otimes Q_{i+1}\otimes\dots\otimes Q_{n}) \rho_j^{\otimes n}) \\
=tr(P_{i}\rho_j) \prod\limits_{k \neq i}{tr(Q_{k}\rho_j)} \\
=0,
\end{array}
\end{equation}
where the second equality holds because the formula of the second
step includes the item $tr(Q_{j}\rho_j)=0$.

Therefore the measurement outcome corresponding to $Q_{1} \otimes
\dots Q_{i-1} \otimes P_{i} \otimes Q_{i+1}\otimes\dots\otimes
Q_{n}$ correctly indicates the state $\rho_i^{\otimes n}$ for any
$i=1,\dots,n$. For any other measurement outcome, we get an
inconclusive result. This is an unambiguous discrimination
strategy among the states $\rho_1^{\otimes
n},\dots,\rho_n^{\otimes n}$. \hfill $\blacksquare$

Applying Lemma \ref{lemma_state}, we find the necessary and
sufficient condition for a set of quantum operations to be
unambiguously distinguishable by multiple uses.

\begin{theorem}
If the quantum operations $\{{\cal E}_1,\dots,{\cal E}_n\}$
satisfy that for any $i \neq j$, $supp({\cal E}_i) \not \subseteq
supp({\cal E}_j)$, then they can be unambiguously discriminated by
$n$ uses. Otherwise, for any $N \ge 1$, they cannot be
unambiguously discriminated by $N$ uses.
\label{theorem_multiple}
\end{theorem}

{\it Proof.} If there exist ${\cal E}_i$ and ${\cal E}_j$ such
that $supp({\cal E}_i) \subseteq supp({\cal E}_j)$, then it is
easy to see that for any $N \ge 1$, $supp({\cal E}_i^{\otimes N})
\subseteq supp({\cal E}_j^{\otimes N})$. It follows from Theorem
\ref{theorem_single} that ${\cal E}_i^{\otimes N}$ and ${\cal
E}_j^{\otimes N}$ cannot be unambiguously discriminated by a
single use.

Now suppose that for any $i \neq j$, $supp({\cal E}_i) \not
\subseteq supp({\cal E}_j)$. From Theorem \ref{theorem_single} we
know that for any $i \neq j$, ${\cal E}_i$ and ${\cal E}_j$ are
unambiguously distinguishable by a single use. Furthermore, it is
from the proof of Theorem \ref{theorem_single} that for any
entangled input $|\psi\rangle$ with full Schmidt number, the
outputs $({\cal E}_i\otimes {\cal I})(|\psi\rangle\langle\psi|)$
and $({\cal E}_j\otimes {\cal I})(|\psi\rangle\langle\psi|)$ are
unambiguously distinguishable. So the set of states $\{({\cal
E}_1\otimes {\cal I})(|\psi\rangle\langle\psi|),\dots,({\cal
E}_n\otimes {\cal I})(|\psi\rangle\langle\psi|)\}$ satisfy the
condition of Lemma \ref{lemma_state}, we hereby conclude that
their $n$-fold copies $(({\cal E}_1\otimes {\cal
I})(|\psi\rangle\langle\psi|))^{\otimes n},\dots, (({\cal
E}_n\otimes {\cal I})(|\psi\rangle\langle\psi|))^{\otimes n}$ can
be unambiguously discriminated. Thus ${\cal E}_1,\dots,{\cal E}_n$
can be unambiguously discriminated by $n$ uses with the input
$|\psi\rangle^{\otimes n}$. \hfill$\blacksquare$

Combining the proof of Lemma \ref{lemma_state} and Theorem
\ref{theorem_multiple}, we have explicitly constructed the input
states and corresponding measurements that unambiguously
discriminate the given quantum operations in the case of multiple
uses. It should be noted that the measurement presented in the
proof Lemma \ref{lemma_state} is actually separable so it is
practically implementable.

Comparing the condition of Theorem \ref{theorem_multiple} with
that of Theorem \ref{theorem_single}, we can see that the former
is looser. So it is possible that a set of quantum operations can
be unambiguously discriminated only by multiple uses. For example,
consider three Pauli channels ${\cal E}_1,{\cal E}_2,{\cal E}_3$
which have Kraus operators $\{\sqrt{p}I,\sqrt{1-p}X\}$,
$\{\sqrt{q}I,\sqrt{1-q}Z\}$ and $\{\sqrt{s}X,\sqrt{1-s}Z\}$
respectively, i.e.
\begin{eqnarray}
{\cal E}_1(\rho) & = & p\rho+(1-p)X\rho X, \\
{\cal E}_2(\rho) & = & q\rho+(1-q)Z\rho Z, \\
{\cal E}_3(\rho) & = & sX\rho X +(1-s) Z \rho Z.
\end{eqnarray}
Since $supp({\cal E}_3)\subseteq supp(\{{\cal E}_1,{\cal E}_2\})$,
we know from Theorem \ref{theorem_single} that they cannot be
unambiguously discriminated by a single use. However, they satisfy
the condition of Theorem \ref{theorem_multiple} and thus can be
unambiguously discriminated by three uses.

It is also found that when distinguishing two quantum operations,
the conditions of Theorem \ref{theorem_single} and Theorem
\ref{theorem_multiple} coincide, which means that multiple uses do
not change the distinguishability in this case. By Theorem
\ref{theorem_multiple}, the only scenario in which unambiguous
discrimination cannot be applied to a given set of quantum
operations is that one of them has support totally contained in
the support of another one.

Our discussions above mainly focus on the possibility of
unambiguous discrimination. It is certainly beneficial to consider
how to achieve the best efficiency. Specifically, we should find
the optimal input state that maximizes the optimal success
probability of unambiguous discrimination between the
corresponding output states. But even for a set of known states,
the optimal success probability of unambiguous discrimination
between them has no analytical formulation in general so far
\cite{ES04}. So our problem is difficult to solve analytically.
Even so, since in most cases the quantum operation to be
identified is repeatable, we can repeat a nonoptimal procedure to
make the total failure probability exponentially decrease,
obtaining the result quickly. Because our strategy is error-free,
once we get a conclusive result it can be immediately accepted.

It is a surprising fact that the conditions for unambiguous
discrimination of quantum operations are in the form similar to
those for quantum states. This can be understood partially from
the point that our strategy has a natural dependence on
unambiguous discrimination of quantum states. A profounder
understanding is that there exist inherent relations between
quantum states and operations. Some previous work has been devoted
to this topic \cite{JA72,CD01,DP03} and it deserves further
research.

In conclusion, we consider the problem of unambiguous
discrimination among a given set of quantum operations. We derive
the necessary and sufficient condition for them to be
unambiguously distinguishable in the cases of single use and
multiple uses respectively. In the latter case a strategy is
explicitly given. It is also found that the use of entanglement
can improve the efficiency and even change the possibility of
unambiguous discrimination between the given quantum operations.
We hope our work can stimulate further research on discrimination
of quantum operations.

We are thankful to the colleagues in the Quantum
Computation and Quantum Information Research Group for helpful
discussions.


\begin{thebibliography}{99}
\bibitem{SS} The support of a density operator is the space
spanned by its eigenvectors with nonzero eigenvalues. The sum of
sets $S_1+\dots+S_n$ is defined to be the set of all the elements
that can be written as $x_1+\dots+x_n$ where $x_i \in S_i$.
\bibitem{K} The kernel of a density operator is the space
orthogonal to its support.
\end{thebibliography}
\end{document}